\documentclass[pre,twocolumn,showpacs]{revtex4}
\usepackage{amsmath}
\usepackage{amssymb}
\usepackage{epsfig}

\begin{document}

\title{Explicit equations for two-dimensional water waves with constant vorticity} 
\author{V.~P. Ruban}
\email{ruban@itp.ac.ru}
\affiliation{Landau Institute for Theoretical Physics,
2 Kosygin Street, 119334 Moscow, Russia} 

\date{\today}

\begin{abstract}
Governing equations for two-dimensional inviscid free-surface flows with
constant vorticity over arbitrary non-uniform bottom profile 
are presented in exact and compact form using conformal variables. 
An efficient and very accurate numerical method for this problem is developed.

\end{abstract}




\maketitle


Most theories for surface water waves have been developed under assumption 
of irrotational flows (see, for example, Refs.\cite{Zakharov67,Z1999,LZ2005},
and references therein). 
The reason is that irrotational incompressible flows are completely 
determined by the boundary distribution of the velocity potential through 
solution of the Laplace equation, and this fact effectively reduces 
spatial dimensionality of the problem.
However, in many real situations waves propagate on shear currents, and 
over non-uniform depth. Interaction between waves and currents is important in
many aspects, for example,  as a mechanism of giant wave formation
\cite{Peregrine,Kharif-Pelinovsky,LP2006}.
The fully nonlinear problem of vortical flows under free surface 
is very complicated. Various simplifications are assumed in theoretical studies,
as weak nonlinearity, mild slope, long wave length, and so on (see, 
for examples, Refs.\cite{B1962,FJ1970,Y1972,J1990,Kirby,B2007}).
The only possibility for non-potential flows, when dimensionality is reduced, 
is the two-dimensional (2D) problem with constant vorticity,
since in this case all perturbations are irrotational. In particular,
relatively long solitary waves in water of finite uniform depth 
with constant vorticity were calculated analytically 
\cite{Miroshnikov2002,Choi2003}. There are also fully nonlinear numerical 
results for steady flows obtained by a boundary integral method 
(see Refs.\cite{SilvaPeregrine1988,PullinGrimshaw1988,VdB1994-1996}, 
and references therein).
At the same time, unsteady rotational waves have not been 
extensively simulated. The present work is intended to fill this gap.
Here for the first time an exact and compact formulation of the fully 
nonlinear problem of waves with constant vorticity  is presented in terms 
of so called conformal variables. The conformal variables
were introduced earlier and successfully used 
to describe purely potential flows on a constant depth or on 
infinitely deep water \cite{DKSZ96,DZK96,ZDV2002,DZ2005Pisma,ZDP2006}. 
Later, the description has been generalized by the present author
to the case of potential flows over arbitrary non-uniform and 
time-dependent bottom profile \cite{R2004PRE,R2005PLA}. 
One of the main advantages of equations in
conformal variables is an easy numerical implementation with fast Fourier
transform (FFT) subroutines. In this work, after derivation of exact
evolutionary equations, some illustrative numerical results will be presented.
Here, for simplicity, a non-uniform bottom profile  does not 
depend on time, though a generalization to non-static bottom is straightforward.


{\bf Derivation of exact equations}.
In what follows, style and notations will be the same as in 
Ref.\cite{R2005PLA}. 
We consider here a 2D incompressible inviscid non-stationary
flow in $(x,y)$-plane, bounded by a given bottom profile from below 
and by an unknown free surface from above.
The flow is rotational, with a constant vorticity field,
$\partial_x V_{(y)}-\partial_y V_{(x)}=-\Omega=const.$
The velocity field thus can be represented as follows,
${\bf V}=(\Omega y+\varphi_x,\quad \varphi_y),$
where the potential $\varphi(x,y,t)$ satisfies the Laplace equation
$\varphi_{xx}+\varphi_{yy}=0$.
Let us also introduce a harmonically conjugate function $\theta(x,y,t)$:
$\varphi_x=\theta_y,\quad \varphi_y=-\theta_x.$
Then it is easy to check that two components of the vector Euler equation are
equivalent to a single scalar equation, analogous to the Bernoulli equation,
\begin{equation}\label{Bernoulli_generalized}
\varphi_t-\Omega\theta+\Omega y\varphi_x +
({\varphi_x^2+\varphi_y^2})/{2} + gy + {p}/{\rho_*}=const,
\end{equation}
where $g$ is the gravity acceleration, $\rho_*$ is a constant density of the 
fluid, and $p(x,y,t)$ is the pressure. It is Eq.(\ref{Bernoulli_generalized}) 
that makes possible reduction of dimensionality of the problem, 
in the same manner as for purely potential flows.

Important point is that complex combination 
$\tilde\phi(z,t)=\varphi(x,y,t)+i\theta(x,y,t)$ is an analytic function 
of the complex argument $z=x+iy$. Analyticity is preserved under conformal
coordinate transforms $z=z(w,t)$, where $w=u+iv$ is a new complex variable. 
We choose an analytic function  $z(w,t)$ in such a manner that 
$w=u$ at the bottom and $w=u+i$ at the free surface. Shape of the surface will
be given in a parametric form,
\begin{equation}\label{surface_shape}
X^{(s)}(u,t)+iY^{(s)}(u,t)\equiv Z^{(s)}(u,t)=z(u+i,t),
\end{equation}
The bottom profile will be determined by
\begin{equation}\label{bottom_shape}
X^{(b)}(u,t)+iY^{(b)}(u,t)\equiv  Z^{(b)}(u,t)=z(u,t).
\end{equation}
Thus, we have an analytic function  $\phi(w,t)=\tilde\phi(z(w,t),t)$ 
defined in the stripe $0\le v\le 1$ in the $(u,v)$-plane. 
Let us designate boundary values of this function as written below,
\begin{equation}\label{boundary_potential}
\phi(u+i,t)\equiv \Phi^{(s)}(u,t),\qquad \phi(u,t)\equiv \Phi^{(b)}(u,t).
\end{equation}
Since $\Phi^{(s)}(u,t)$ and $\Phi^{(b)}(u,t)$ are values of the same analytic
function at points $u$ and $u+i$, 
they are related to each other by a linear transform (see \cite{R2005PLA}),
\begin{equation}\label{Phi_s_Phi_b_relation}
\Phi^{(s)}(u,t)=e^{-\hat k}\Phi^{(b)}(u,t),
\end{equation}
with $e^{-\hat k}\equiv\exp(i\hat\partial_u)$. That means 
$\Phi^{(s)}_k(t)=e^{-k}\Phi^{(b)}_k(t)$ for the corresponding Fourier images.

The velocity components are determined by the following relations
\begin{eqnarray}\label{VxVy}
V_{(x)}-iV_{(y)}&=&\Omega y+\varphi_x-i\varphi_y
=\Omega\,\mbox{Im}z+d\tilde \phi/dz \nonumber\\
&=&\Omega\,\mbox{Im}z(w,t)+\phi'(w,t)/z'(w,t).
\end{eqnarray} 

Now we are going to write equations of motion in the conformal variables.
First, we have two kinematic conditions which in our case take form
\begin{eqnarray}\label{kinematic_Dw_s}
-\mbox{Im}\left(Z^{(s)}_t\bar Z^{(s)}_u\right)&=&
[\mbox{Im\,}\Phi^{(s)}+(\Omega/2)(\mbox{Im\,}Z^{(s)})^2]_u,\\
\label{kinematic_Dw_b}
0&=&[\mbox{Im\,}\Phi^{(b)}+(\Omega/2)(\mbox{Im\,}Z^{(b)})^2]_u,
\end{eqnarray}
where $\bar Z$ denotes complex conjugate value, and the subscripts denote 
the corresponding partial derivatives. At free surface the pressure is
constant (we neglect here surface tension $\sigma$, otherwise 
$p^{(s)}=\sigma\kappa+const$, where $\kappa$ is the surface curvature).
Therefore from Eq.(\ref{Bernoulli_generalized}) we have the dynamic boundary
condition in conformal variables,
\begin{eqnarray}
&&\mbox{Re}\left(\Phi^{(s)}_t-
\Phi^{(s)}_u{Z^{(s)}_t}/{Z^{(s)}_u}\right) 
+|\Phi^{(s)}_u/Z^{(s)}_u|^2/2+g\,\mbox{Im\,}Z^{(s)}\nonumber\\
\label{Bernoulli_Dw}
&&-\,\Omega\,\mbox{Im\,}\Phi^{(s)}+
\Omega\,\mbox{Im\,}Z^{(s)}\mbox{Re}\left(
{\Phi^{(s)}_u}/{Z^{(s)}_u}\right) =0.
\end{eqnarray}
Taking into account Eq.(\ref{kinematic_Dw_b}),
it is convenient to represent $\Phi^{(b)}(u,t)$ in the form
\begin{equation}\label{Phi_b}
\Phi^{(b)}=\hat S \psi 
-i(\Omega/2) (1-i\hat R)[Y^{(b)}]^2,
\end{equation}
where $\psi(u,t)$ is some unknown real function, and the linear operators 
$\hat S$ and $\hat R$ are diagonal in Fourier representation:
$ S_k={1}/{\cosh(k)},\quad  R_k=i\tanh(k).$
It is essential that when $\hat S$ or $\hat R$ acts on a purely real function,
the result is also real.
We will also need operator $T_k=-i\coth(k)=R_k^{-1}$. Since 
$e^{-\hat k}\hat S=(1+i\hat R)$ and $e^{-\hat k}(1-i\hat R)=\hat S$,  
we have from Eqs.(\ref{Phi_s_Phi_b_relation}) and (\ref{Phi_b})
the following formula for $\Phi^{(s)}(u,t)$,
\begin{equation}\label{Phi_s}
\Phi^{(s)}=(1+i\hat R)\psi -i(\Omega/2) \hat S [Y^{(b)}]^2.
\end{equation}

Now we should take into account that the function $z(w,t)$ 
can be represented as a composition
of two functions (see \cite{R2004PRE,R2005PLA}), that is
$z(w,t)=Z(\zeta(w,t)),$
where a known analytic function $Z(\zeta)=X(\zeta)+iY(\zeta)$ 
determines bottom shape. The conformal mapping $Z(\zeta)$ does not have 
any singularities within a sufficiently wide horizontal stripe above 
the real axis in $\zeta$-plane.
An intermediate analytic function $\zeta(w,t)$ takes real values 
at the real axis, and therefore 
\begin{equation}\label{zeta_def}
\zeta(w,t)=\int \frac{a_k(t)}{\cosh(k)}e^{ikw}\frac{dk}{2\pi}, 
\qquad a_{-k}=\bar a_k,
\end{equation}
where $a_k(t)$ is Fourier transform of a real function $a(u,t)$.
On the bottom $\zeta(u,t)=\hat S a(u,t)$, therefore
\begin{equation} \label{Zb}
Z^{(b)}(u,t)=Z(\hat S a(u,t)), 
\end{equation}
At the free surface we have relations
\begin{equation}\label{xi_def}
\zeta(u+i,t)\equiv\xi(u,t)=(1+i\hat R)a(u,t),
\end{equation}
and
$Z^{(s)}=Z(\xi),\quad Z^{(s)}_u=Z_\xi(\xi)\xi_u, 
\quad Z^{(s)}_t=Z_\xi(\xi)\xi_t.$


Thus we have in our system two unknown real functions, 
$\psi(u,t)$ and $a(u,t)$. 
All the other quantities are expressed through these two. 
Our purpose now is to derive equations determining time derivatives 
$\psi_t$ and $a_t$.
To do this, we divide Eq.(\ref{kinematic_Dw_s}) by $|Z^{(s)}_u|^2$ and obtain
that
$\mbox{Im}({\xi_t}/{\xi_u})=-Q$,
where
\begin{equation}\label{Q_def}
Q\equiv\frac{\{\hat R \psi +(\Omega/2)([\mbox{Im\,}Z(\xi)]^2
-\hat S[\mbox{Im\,}Z(\hat S a)]^2)\}_u}
{|Z_\xi(\xi)\xi_u|^2}.
\end{equation}
Since  $\xi_t/\xi_u=\zeta_t(w,t)/\zeta_w(w,t)|_{w=u+i}$, 
there exists a relation 
between the real and imaginary parts: 
$\mbox{Im}(\xi_t/\xi_u)=\hat R\,\mbox{Re}(\xi_t/\xi_u)$,
so $\mbox{Im}({\xi_t}/{\xi_u})=-Q$ means $\xi_t=-\xi_u(\hat T+i)Q$ and it 
gives us equation determining $a_t$,
\begin{equation}\label{a_t} 
a_t=-\mbox{Re}[\xi_u(\hat T+i)Q].
\end{equation}
After that, Eqs.(\ref{Bernoulli_Dw}) and (\ref{Phi_s}) allow us to express
$\psi_t$:
\begin{eqnarray}
\psi_t&=&-\mbox{Re}[\Phi^{(s)}_u(\hat T+i)Q] 
-|\Phi^{(s)}_u/Z^{(s)}_u|^2/2
-g\,\mbox{Im\,}Z^{(s)}\nonumber\\
&+&\Omega\,\mbox{Im\,}\Phi^{(s)}
-\Omega\,\mbox{Im\,}Z^{(s)}\mbox{Re}\left(
{\Phi^{(s)}_u}/{Z^{(s)}_u}\right).
\label{psi_t}
\end{eqnarray}
Now, exact and explicit evolutionary equations have been derived. In the next
section it is explained how one can numerically simulate them with a high 
accuracy.


\begin{figure}
\begin{center}
\epsfig{file=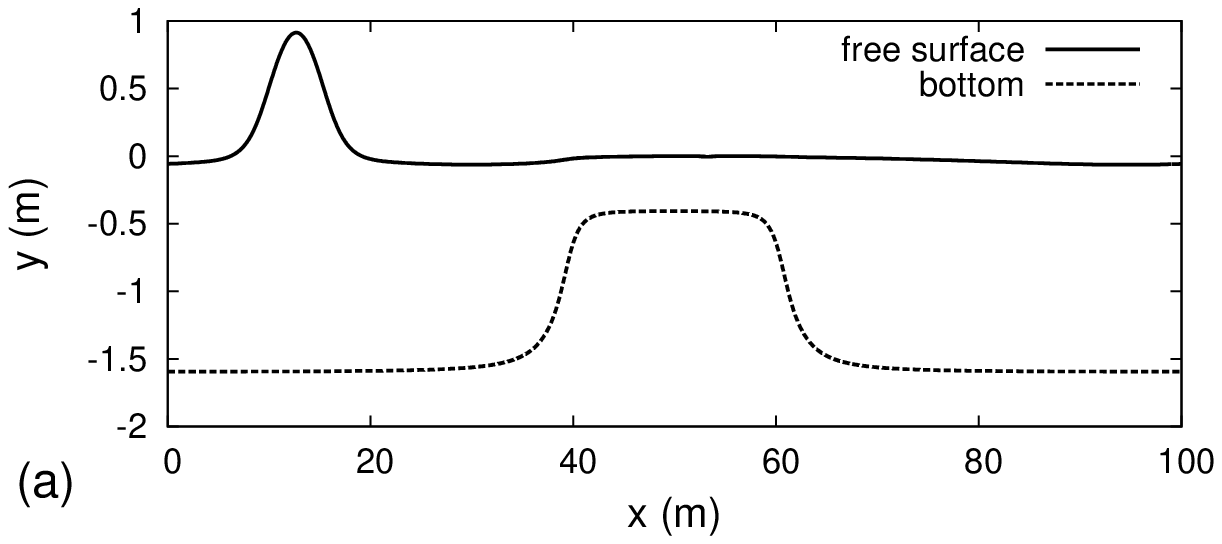,width=86mm}\\
\epsfig{file=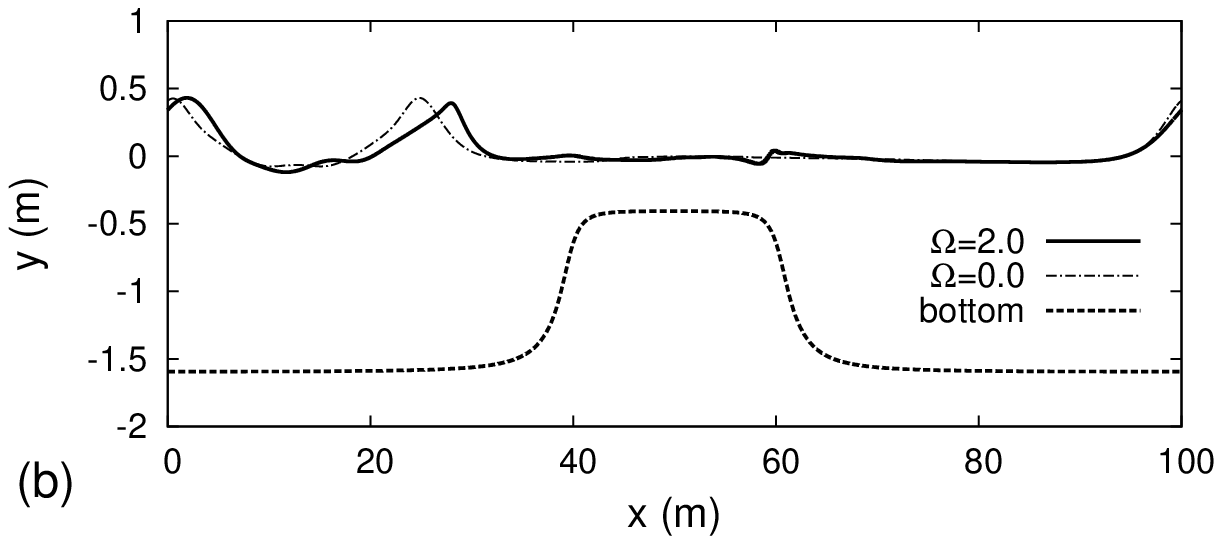,width=86mm}\\
\epsfig{file=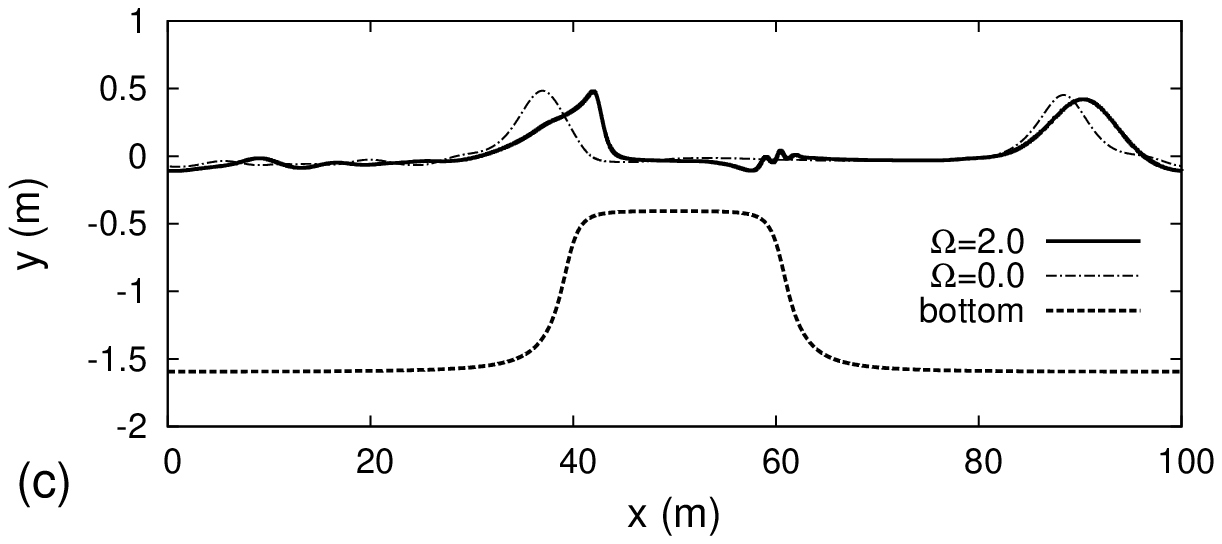,width=86mm}\\
\epsfig{file=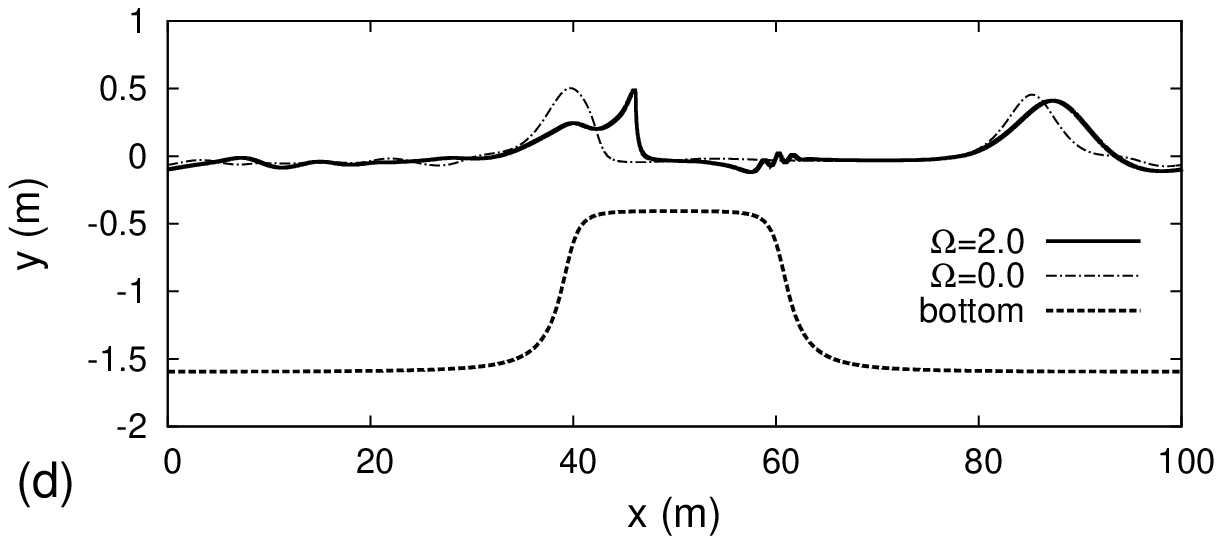,width=86mm}
\end{center}
\caption{Shape of the free surface at different time moments;
a) $t=0.00$ s; b) $t=2.8$ s; c) $t=5.6$ s, d) $t=6.32$ s. } 
\label{t_0_1_2_3} 
\end{figure}

\begin{figure}
\begin{center}
\epsfig{file=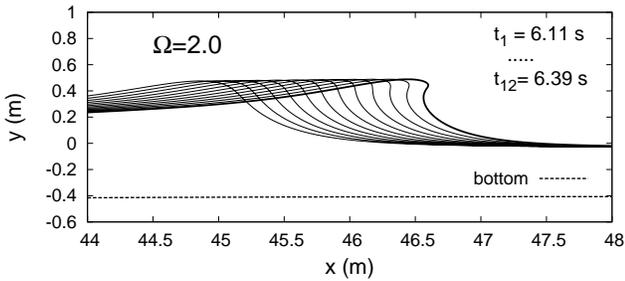,width=86mm}
\end{center}
\caption{Profiles of the overturning wave from $t=6.11$ s (left curve) 
to $t=6.39$ s (right curve).} 
\label{overturning_wave} 
\end{figure}

\begin{figure}
\begin{center}
\epsfig{file=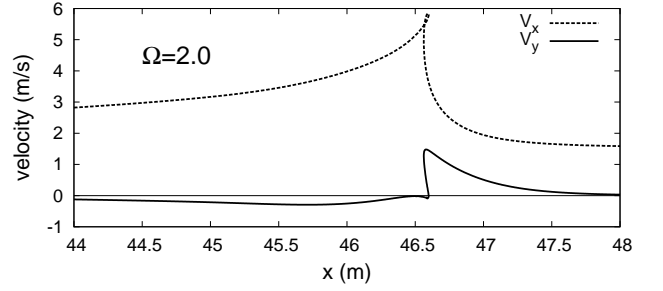,width=86mm}
\end{center}
\caption{Velocity distribution at the free surface for $t=6.39$s.} 
\label{velocity} 
\end{figure}

{\bf Numerical method and example}.
Let us consider the case when a bottom profile is periodic on $x$.
Obviously, there exist solutions with velocity field and surface
elevation having the same spatial period $L$. Without loss of generality, 
the potential $\varphi(x,y,t)$ can be assumed periodic (the part
of $\varphi$ which is proportional to $x$ and corresponds to a constant 
horizontal velocity, can be excluded by a redefinition of $y$ zero level).
Making a proper choice for the length and time scales, we may write $g=1$, 
and $Z(\zeta+2\pi)=2\pi+Z(\zeta)$.  
Direct substitution into 
Eqs.(\ref{a_t}) and (\ref{psi_t}) shows the solutions are $2\pi$-periodic
on the variable $\vartheta=u\alpha(t)$, where a real function $\alpha(t)$ 
depends on time in a non-trivial manner 
in order to cancel non-periodic terms in Eqs.(\ref{a_t}) and (\ref{psi_t}).
The non-periodic terms take place because the operator $\hat T$ is singular 
at small $k$, and its action on a constant function $Q_0$ is non-periodic:
$\hat T Q_0=Q_0u$ (see Refs.\cite{R2004PRE,R2005PLA}). 
So, we can write for rescaled dimensionless quantities
\begin{eqnarray}
a(\vartheta,t)&=&\vartheta+
\sum_{m=-\infty}^{+\infty}\rho_m(t)\exp(im\vartheta),
\label{rho_m}
\\
\xi(\vartheta,t)&=&\vartheta+i\alpha(t)
+\sum_{m=-\infty}^{+\infty}\frac{2\rho_m(t)\exp(im\vartheta)}
{1+\exp(2m\alpha(t))}, 
\label{xi_m} 
\\
\psi(\vartheta,t)&=&\sum_{m=-\infty}^{+\infty}\psi_m(t)\exp(im\vartheta),
\label{psi_m}
\end{eqnarray}
where $\rho_m(t)$ and $\psi_m(t)$ are Fourier coefficients of 
$2\pi$-periodic real functions $\rho(\vartheta,t)$ and $\psi(\vartheta,t)$.
As a result, in variables $(\vartheta,t)$ equations for $\rho_t$ 
and $\psi_t$ look similar to Eqs.(\ref{a_t}) and (\ref{psi_t}), 
but all the $u$-derivatives should be replaced by $\vartheta$-derivatives, 
and operators $\hat R$, $\hat S$, and $\hat T$ should be everywhere replaced 
by new operators $\hat{\mathsf R}_\alpha$, $\hat{\mathsf S}_\alpha$, and
$\hat{\mathsf T}_\alpha$ respectively:
\begin{eqnarray}
\rho_t&=&-\mbox{Re}[\xi_\vartheta(\hat {\mathsf T}_\alpha+i){\mathsf Q}],
\label{rho_t_alpha} 
\\
\psi_t&=&-\mbox{Re}
[\Phi_\vartheta(\hat{\mathsf T}_\alpha +i){\mathsf Q}] 
-\frac{|\Phi_\vartheta|^2}{2|Z_\xi(\xi)\xi_\vartheta|^2}
-g\,\mbox{Im\,}Z(\xi)\nonumber\\
&&+\,\Omega\,\mbox{Im\,}\Phi
-\Omega\,\mbox{Im\,}Z(\xi)\,\mbox{Re}\left(
\frac{\Phi_\vartheta}{Z_\xi(\xi)\xi_\vartheta}\right),
\label{psi_t_alpha}
\end{eqnarray}
where $\xi=\vartheta+i\alpha+(1+i\hat{\mathsf R}_\alpha)\rho$, and
\begin{eqnarray}
\Phi&=&(1+i\hat{\mathsf R}_\alpha)\psi 
-i(\Omega/2) \hat{\mathsf S}_\alpha 
Y^2(\vartheta+\hat{\mathsf S}_\alpha \rho),\\
{\mathsf Q}&=&\frac{\{\hat{\mathsf R}_\alpha \psi 
+(\Omega/2)[Y^2(\xi)
-\hat{\mathsf S}_\alpha
Y^2(\vartheta+\hat{\mathsf S}_\alpha \rho)]\}_\vartheta}
{|Z_\xi(\xi)\xi_\vartheta|^2}.
\end{eqnarray}
These new operators are diagonal in the discrete Fourier representation: 
${\mathsf R}_\alpha(m)=i\tanh(\alpha m)$,
${\mathsf S}_\alpha(m)=1/\cosh(\alpha m)$, and
${\mathsf T}_\alpha(m)=-i\coth(\alpha m)$ for $m\not=0$,
${\mathsf T}_\alpha(0)=0$.
The system of equations is closed by the following condition for 
$\dot\alpha(t)$, which ensures cancellation of the non-periodic terms 
in Eqs.(\ref{a_t}) and (\ref{psi_t}),
\begin{equation}\label{dot_alpha}
\dot\alpha(t)=-\frac{1}{2\pi}\int_0^{2\pi}{\mathsf Q}(\vartheta)d\vartheta.
\end{equation}
The above system of equations has two apparent
integrals of motion, namely the area $A$
occupied by fluid, 
$A=A_0+\int Y(\xi)\,\mbox{Re}[Z_\xi(\xi)\xi_\vartheta]\,d\vartheta,$
and the total energy $E$ (kinetic energy plus potential energy in the
gravitational field),
\begin{eqnarray}
E&=&E_0+\frac{\Omega^2}{6}\int Y^3(\xi)\,
\mbox{Re}[Z_\xi(\xi)\xi_\vartheta]\,d\vartheta\nonumber\\
&+&\frac{\Omega^2}{8}\int Y^2(\vartheta+\hat{\mathsf S}_\alpha \rho)
\hat{\mathsf R}_\alpha [Y^2(\vartheta+\hat{\mathsf S}_\alpha \rho)]_\vartheta
\,d\vartheta\nonumber\\
&+&\frac{\Omega}{2}\int[Y^2(\xi)-\hat{\mathsf S}_\alpha
Y^2(\vartheta+\hat{\mathsf S}_\alpha \rho)]
\psi_\vartheta\,d\vartheta\nonumber\\
&\!-\!&\frac{1}{2}\!\int\!\psi\,\hat{\mathsf R}_\alpha\psi_\vartheta\,
d\vartheta
+\frac{g}{2}\!\int\! Y^2(\xi)\,\mbox{Re}[Z_\xi(\xi)\xi_\vartheta]
\,d\vartheta,
\end{eqnarray}
where $A_0$ ans $E_0$ are constant, 
and all the integrals are in the limits from $0$ to $2\pi$.

Eqs.(\ref{rho_t_alpha})-(\ref{dot_alpha}) are easy for numerical simulation
if the function $Z(\zeta)$ is given by a simple formula as it takes
place for many interesting bottom profiles.
The numerical method employed here is naturally based on the discrete Fourier
representation, since all the linear operators in the equations
are efficiently computed with modern FFT subroutines in $m$-representation, 
while all the nonlinear operations are simple in $\vartheta$-representation.
As primary dynamical variables, the quantities $\alpha(t)$, $\rho_m(t)$, 
and $\psi_m(t)$ are taken, with $0\le m < M$ 
(for negative $m$ the relations $\rho_{-m}=\bar\rho_m$ and 
$\psi_{-m}=\bar\psi_m$ are used). After each step of a Runge-Kutta 4-th 
order procedure, only spectral components with $|m|< M_{eff}$ are kept, where
$M_{eff}\approx (1/4)N$, $M\approx (3/8)N$, and $N=2^{12...19}$ 
is the size of arrays for the fast Fourier transform (during computations, 
$N$ is doubled several times as small-scale structures develop). 
As a result of the adaptive increasing of $N$, 
the right hand sides of Eqs.(\ref{rho_t_alpha})-(\ref{psi_t_alpha}) 
can be computed with nearly the same numerical error $\delta_0< N 10^{-18}$ 
as it is for the FFT subroutine using C-type {\tt double}. 
Since the time step is decreased as $\tau\sim 1/N$ for the stability reasons,
an error for the free surface position at $t\sim 1$ can be estimated as 
$\delta\lesssim N^2 10^{-18}$.
Practically, $A$ and $E$ are conserved up to $10$ decimal digits
for most part of the evolution. 
In a final stage, the larger $N_{final}$ is used, the later time moment is
when the high accuracy is lost.

Here an example is given which demonstrates  potentialities of the method. 
Let the bottom profile be determined by formula 
$Z(\zeta)=iY_0+{\cal B}(\zeta-i\alpha_0)$, where $Y_0=0.02\pi$, 
$\alpha_0=0.02\pi$, and
$$
{\cal B}(q)=q-i\Delta\mbox{Ln}\left[
\left(i\sin q+\sqrt{\epsilon +\cos^2 q}\right)
(1+\epsilon)^{-1/2}\right],
$$
with $\Delta=0.6$ and  $\epsilon=0.02$. At $t=0$ we put $\alpha=\alpha_0$ and 
$
\rho=0.05\tanh[15\sin(\vartheta-\gamma)]
\exp[-2(1-\cos(\vartheta-\gamma))],
$ 
where $\gamma=0.5$.
In Fig.\ref{t_0_1_2_3}-a, the bottom profile and the initial surface shape 
are presented normalized to the spatial period $L=100$ m and shifted
appropriately in the vertical direction.
Dimensionless vorticity $\Omega=2.0$ together with the parameter $Y_0$
give the rotational part of the horizontal velocity field 
$V^{(\Omega)}_{(x)}\approx 1.6(y+1)$ m/s. This corresponds to a backward flow 
along the bottom in the deeper regions. 
We choose $\psi(\vartheta,0)$ in such a way that at $t=0$  the normal component 
of the total velocity field at the free surface is zero:
$
\psi(\vartheta,0)=-({\Omega}/{2})\hat{\mathsf T}_\alpha[Y^2(\xi)
-\hat{\mathsf S}_\alpha Y^2(\vartheta+\hat{\mathsf S}_\alpha \rho)].
$
Some results of the computation are presented in 
Fig.\ref{t_0_1_2_3}, where also a comparison is made
to the case $\Omega=0$.
The initial hump at the surface decays into two oppositely propagating
solitary waves having different speeds and different
profiles (crest of the right-propagating wave is more sharp; 
however, a maximum curvature is finite).
In this example the right-propagating wave first meets the region 
of relatively shallow depth,
where its crest becomes more and more steep, and finally the wave profile 
overturns, as it is seen in Fig.\ref{overturning_wave}. 
Velocity distribution along the overturning wave is shown 
in  Fig.\ref{velocity}.
Another interesting phenomenon
observed here is the wave blocking near $x\approx 60$ m, where 
an average horizontal flow velocity ($\sim 1.6$ m/s for $\Omega=2.0$) 
approaches speed of typical waves $v_{ph}\approx\sqrt{gh}$ 
($h$ is the local depth),
and therefore waves cannot enter the shallow region from the right.
In different simulations,
for $\Omega\gtrapprox 2.4$, the blocking was so strong that wave height
near the point $x=60$ was comparable to the local depth (not shown). 
However, an extensive discussion of this phenomenon is not possible 
in this Brief Report.

{\bf Acknowledgments}.These investigations were supported 
by RFBR Grant 06-01-00665, by the Program ``Fundamental Problems of 
Nonlinear Dynamics'' from the RAS Presidium, 
and by Grant ``Leading Scientific Schools of Russia''.


\begin{thebibliography}{99}

\bibitem{Zakharov67} V.E. Zakharov, Sov. Phys. JETP {\bf 24}, 455 (1967).

\bibitem{Z1999} V. E. Zakharov, Eur. J.  Mech. B/Fluids {\bf 18}, 327 (1999).

\bibitem{LZ2005} P. M. Lushnikov and V. E. Zakharov, 
Physica D {\bf 203}, 9 (2005).

\bibitem{Peregrine} D.H. Peregrine, Adv. Appl. Mech. {\bf 16}, 9 (1976).

\bibitem{Kharif-Pelinovsky} C. Kharif and E. Pelinovsky,
Eur. J. Mech. B/Fluids {\bf 22}, 603 (2003). 

\bibitem{LP2006} I.V. Lavrenov and A.V. Porubov,
Eur. J. Mech. B/Fluids {\bf 25}, 574 (2006).

\bibitem{B1962} T.B. Benjamin, J. Fluid Mech. {\bf 12}, 97 (1962).

\bibitem{FJ1970} N.C. Freeman and R.S. Johnson, 
J. Fluid Mech. {\bf 42}, 401 (1970).

\bibitem{Y1972} C.S. Yih, J. Fluid Mech. {\bf 51}, 209 (1972).

\bibitem{J1990} R.S. Johnson, J. Fluid Mech. {\bf 215}, 145 (1990).

\bibitem{Kirby} {\tt http://chinacat.coastal.udel.edu/\~{}kirby/}

\bibitem{B2007} K.A. Belibassakis, J. Fluid Mech. {\bf 578}, 413 (2007).


\bibitem{Miroshnikov2002} V. A. Miroshnikov, 
J. Fluid Mech. {\bf 456}, 1 (2002).

\bibitem{Choi2003} W. Choi, Phys. Rev. E {\bf 68}, 026305 (2003).

\bibitem{SilvaPeregrine1988} A.F. Teles da Silva and D. H. Peregrine,
J. Fluid Mech. {\bf 195}, 281 (1988).

\bibitem{PullinGrimshaw1988} D. I. Pullin and R. H. J. Grimshaw,
Phys Fluids {\bf 31}, 3550 (1988). 

\bibitem{VdB1994-1996} J.-M. Vanden-Broeck, 
J. Fluid Mech. {\bf 274}, 339 (1994);
 J.-M. Vanden-Broeck, Eur. J. Mech. B/Fluids {\bf 14}, 761 (1995);
 J.-M. Vanden-Broeck, IMA J. Appl. Math. {\bf 56}, 207 (1996).

\bibitem{DKSZ96}A. I. Dyachenko {\it et al.}, 
Phys. Lett. A {\bf 221}, 73 (1996).

\bibitem{DZK96} A. I. Dyachenko, V. E. Zakharov, and E. A. Kuznetsov,
Fiz. Plazmy {\bf 22}, 916 (1996) [Plasma Phys. Rep. {\bf 22}, 829 (1996)].


\bibitem{ZDV2002} V. E. Zakharov, A. I. Dyachenko, and O. A. Vasilyev,
Eur. J. Mech. B/Fluids {\bf 21}, 283 (2002).

\bibitem{DZ2005Pisma} A. I. Dyachenko and V. E. Zakharov, 
Pis'ma v ZhETF {\bf 81}, 318 (2005) [JETP Letters {\bf 81}, 255 (2005)].

\bibitem{ZDP2006}V.E. Zakharov, A.I. Dyachenko and A.O. Prokofiev,
Eur. J. Mech. B/Fluids {\bf 25}, 677 (2006).

\bibitem{R2004PRE} V. P. Ruban, Phys. Rev. E {\bf 70}, 066302 (2004).

\bibitem{R2005PLA} V. P. Ruban, Phys. Lett. A {\bf 340}, 194 (2005).

\end{thebibliography}
\end{document}